# WHERE IS THE DARK MATTER?

Neta A. Bahcall, Lori M. Lubin, & Victoria Dorman

Princeton University Observatory, Princeton, NJ 08544

## Abstract

How much dark matter is there in the universe and where is it located? These are two of the most fundamental questions in cosmology. We use in this paper optical and x-ray mass determinations of galaxies, groups, and clusters of galaxies to suggest that most of the dark matter may reside in very large halos around galaxies, typically extending to ~ 200 kpc for bright galaxies. We show that the mass-to-light ratio of galaxy systems does not increase significantly with linear scale beyond the very large halos suggested for individual galaxies. Rather, the total mass of large scale systems such as groups and rich clusters of galaxies, even superclusters, can on average be accounted for by the total mass of their member galaxies, including their large halos (which may be stripped-off in the dense cluster environment but still remain in the clusters) plus the mass of the hot intracluster gas. This conclusion also suggests that we may live in a low-density universe with $\Omega \sim 0.2 - 0.3$.

# 1. Introduction

A large amount of unseen "dark matter" exists in halos around luminous galaxies and in groups and rich clusters of galaxies (Ostriker *et al.* 1974; Einasto *et al.* 1974; Faber & Gallagher 1979, hereafter FG; Davis *et al.* 1980; Trimble 1987). The presence of the dark matter component is inferred from the high rotation speed of gas and stars in the outer parts of spiral galaxies (FG; Trimble 1987), the high velocity dispersion (FG; Lauer 1985; Trimble 1987) and extended x-ray halo of hot gas (FG; Forman *et al.* 1985; Trimble 1987; Fabbiano 1989; Mushotzky *et al.* 1994) in elliptical galaxies, and the high velocity dispersion and gas temperature in clusters of galaxies (FG; Trimble 1987; Lubin & Bahcall 1993). The relative contribution of the dark matter component is usually specified in terms of the mass-to-light ratio, M/L; this reflects the total amount of mass relative to the total light within a given scale. It is well known (Ostriker *et al.* 1974; Davis *et al.* 1980; Blumenthal *et al.* 1984; Rubin 1993) that, on average, the M/L ratio increases from the bright, luminous parts of galaxies to their faint halos, with further increase to systems with larger scales such as groups and rich clusters of galaxies. This increase suggests that there is relatively more dark than luminous matter with increasing scale (cf. Rubin 1993). This has led to the general belief that clusters have more dark matter per unit luminosity than individual galaxies and that superclusters may have even more. This widely accepted monotonic increase of M/L with scale determines to a large extent the prevalent views about the location of the dark matter and the total mass density of the universe. The universal mass density can be estimated from the observed mean luminosity density in the universe when multiplied by the M/L ratio observed on large scales. On the scale of clusters (~ 1 Mpc), this method suggests a total mass density of $\Omega \sim 0.2$. If M/L increases with scale, however, the closure density ($\Omega = 1$) may eventually be reached; in this case, the dark matter component on large scales is distributed more diffusely than the luminous galaxies.

In this paper we show that this widely-believed scenario of an ever-increasing M/L is not suggested by observations.

## 2.  Analysis and Results

We examine the observed M/L ratio as a function of scale for spiral and elliptical galaxies, groups, rich clusters, and superclusters of galaxies. The M/L of spirals and ellipticals are determined from the virial mass based on the velocities of stars or gas in the galaxies within a given radius (FG; Bosma 1981; Lauer 1985; Trimble 1987 and references therein). To extend the estimated galaxy mass beyond the luminous or gas-dominated regimes, we use mass estimates derived from motions of faint satellites at larger distances around galaxies (Bahcall & Tremaine 1981; Zaritzky *et al.* 1993), thus probing the dark galaxy halos. We also use the extended x-ray halos observed around nearly isolated ellipticals to estimate their extended mass assuming that the x-ray gas is gravitationally bound to the galaxy (FG; Forman *et al.* 1985; Trimble 1987; Fabbiano 1989; Mushotzky 1994). Recent observations with the ASCA x-ray satellite further support these estimates (Mushotzky *et al.* 1994; Kim & Fabbiano 1995). The estimated dynamical mass  of our own Milky Way Galaxy within ~100 kpc is also used (Kulessa & Lynden-Bell 1992). All measurements are converted to a consistent system of units; we use a Hubble constant of Ho = 100h km s$^{-1}$ Mpc$^{-1}$, and all luminosities, $L_B$, refer to the *total* luminosity of the galaxies in the blue band, corrected for Galactic and internal extinction and redshift.

In Figure 1 we present the observed M/$L_B$ ratio as a function of radius for a sample of bright spiral and elliptical galaxies (above references). The lines connecting the data points indicate measurements at different radii *of the same galaxy*. The large boxes represent the ~ 1σ range of the observed M/$L_B$ at the Holmberg radius for a large sample of bright ellipticals and spirals (FG; Bosma 1981; Lauer 1985; Trimble 1987). As expected, M/$L_B$ increases with radius to scales beyond the luminous part  of the galaxies (typically R $\lesssim$ 20 kpc); this increase reflects the well-established inference of *dark galaxy halos*. The maximum extent observed so far for an individual galaxy halo is R ~ 200 kpc (Bahcall & Tremaine 1981; Kulessa & Lynden-Bell 1992; Zaritzky *et al.* 1993). Most of the elliptical galaxy data on large scales ($\gtrsim$ 20 kpc) is based on the extended x-ray halos observed around these galaxies (references above).

The functions $M/L_B$ (R) that best fit the observed values in Fig. 1 are: $M/L_B$ (spirals) $\simeq$ 60±10 (R Mpc/0.1) $M_\odot/L_{B\odot}$, and $M/L_B$ (ellipticals) $\simeq$ 200±50 (R Mpc/0.1) $M_\odot/L_{B\odot}$. The average $M/L_B$ ratio of ellipticals appears to be larger than that of spirals at the same radius by a factor of approximately three or four. This difference could be due to a relatively higher mass for average ellipticals than for spirals, or a lower blue luminosity due to the absence of bright young blue stars in the old ellipticals. The average blue luminosities of typical ($\sim L^*$) spirals and ellipticals, however, are observed to be comparable, both in the field and in clusters (Binggeli *et al.* 1988); Fig. 1 thus suggests that, on average, ellipticals are more massive than spirals for the *same* $L_B$ and radius.

If the dark halos around typical bright spirals and ellipticals extend to $\sim$ 150-200$h^{-1}$ kpc, as suggested by Fig. 1, the implied total $M/L_B$ ratio of the galaxies is:

$$M/L_B \text{ (spirals, } \lesssim 0.2h^{-1} \text{ Mpc)} \simeq 100h \quad ,$$

$$M/L_B \text{ (ellipticals, } \lesssim 0.2h^{-1} \text{ Mpc)} \simeq 400h \quad .$$

The large value of $M/L_B$ implied for elliptical galaxies in this scenario can thus contribute significantly to the observed increase of $M/L_B$ in rich clusters of galaxies (see below) since clusters are preferentially populated by elliptical and SO galaxies.

A principal new element suggested in this paper is that the average dark halo of ellipticals is more massive than that of spirals of comparable luminosity and has a similar dependence on scale. This picture will be tested as new data emerge.

To compare the $M/L_B$ ratio of single galaxies with that of larger systems such as groups and rich clusters, we use data from different catalogs and mass estimates from several methods. For groups of galaxies we use the CfA group catalog (Ramella *et al.* 1989), the Hickson Compact Group (HCG) catalog (Hickson *et al.* 1992), Morgan groups (Morgan *et al.* 1975), and several groups recently observed in x-rays by ROSAT and ASCA (Ponman & Bertram 1993; Mulchaey *et al.* 1993, 1995; Pildis *et al.* 1995). The group radii range from R $\lesssim$ 0.1 to $\sim$ 1$h^{-1}$ Mpc. For the large catalogs (such as CfA and HCG, each containing 92 groups), the *median* M/L$_B$ values, binned by radius, are used. The group masses within a given radius are based on the virial mass estimate using the galaxy velocity dispersion and/or the group x-ray temperature observations (see above references). The estimated M/L$_B$ for our local group (Trimble 1987) is also used. The results are presented in Figure 2. The error-bars represent typical 1σ uncertainties of the individual x-ray mass determinations and 1σ scatter around the median optical values.

The M/L$_B$ ratios of rich clusters of galaxies, at the scale of R $\simeq$ 1.5h$^{-1}$ Mpc, are also presented in Figure 2. Rich clusters for which more than 20 galaxy redshifts have been measured are included (Struble & Rood 1991; Lubin & Bahcall 1993). The total sample includes 85 rich clusters; the median M/L$_B$ values for several luminosity bins are presented. The virial mass of each cluster is determined within 1.5h$^{-1}$ Mpc radius for an isothermal distribution using the galaxy velocity dispersion and/or x-ray temperature observations; the optical mass determinations are generally consistent with the x-ray mass determinations for the same clusters (Hughes 1989; Lubin & Bahcall 1993). The total cluster luminosity within radius R is taken from the data when available (Oemler 1974; Postman *et al*. 1988; West *et al*. 1989), or otherwise approximated from the Schechter luminosity function of galaxies in clusters and the cluster richness. All luminosities are corrected to our standard system of luminosity.

Figure 2 is a composite of M/L$_B$ as a function of scale for individual galaxies, groups, and rich clusters. The data points represent median values of larger samples, as well as some individual cases. At 1h$^{-1}$ Mpc we also present, for comparison, the independent Ω determination from the cosmic virial theorem (Davis & Peebles 1983a), utilizing galaxy pairwise velocities at r $\lesssim$ 1h$^{-1}$ Mpc. This result is consistent with the M/L$_B$ function. On much larger scales of R ~ 10-20h$^{-1}$ Mpc, we present approximate upper limits to M/L$_B$ for two large superclusters, Shapley and Corona Borealis, for which an upper limit to the velocity dispersion within the superclusters is available: σ$_r$ $\lesssim$ 1287 km s$^{-1}$ for Corona Borealis (Postman *et al*. 1988) and σ$_r$ $\lesssim$ 1230 km s$^{-1}$ for Shapley [from the observed velocity spread around the Hubble line (Raychaudbury *et al* 1991; see also Bahcall *et al*, 1995)]. The $M/L_B$ value for each supercluster reflects the *binding mass* (=1/2 virial mass) of the supercluster for the above velocity dispersion limit. (The superclusters are unlikely to be virialized since the dynamical time is comparable to the Hubble time; the limits assume that they are bound. If the superclusters are not bound, these limits do not apply). Also shown, for comparison, is the *range* of $\Omega$ values obtained on this scale from the Virgocentric infall data, assuming that mass follows light (Davis & Peebles 1983b, Peebles 1993, Strauss & Willick 1995 and references therein). This range is consistent with the supercluster data. At still larger scales, we present the $M/L_B$ ratio obtained by Tully *et al* (1994) at $30h^{-1}$ Mpc using the Least Action method; it represents their best-fit $M/L_B$ value assuming mass follows light. Finally, at $\sim 50h^{-1}$ Mpc , we present the *range* of various recent reported constraints of the $\beta = \Omega^{0.6}/b$ parameter obtained from observations of bulk-flows and redshift-space anisotropies (Strauss & Willick 1995 and references therein; all data with uncertainties $\sigma(\beta) \le 0.3$ are included). A bias parameter b = 1 (i.e., mass traces light) is assumed. (For b ≠ 1, the values increase as $b^{1.67}$, i.e., $\le 2$ for b $\le$ 1.5). The range of reported $\beta$ values (and thus $\Omega$) on these scales is large, covering the range $\Omega \sim 0.2$ to $\sim 1$, with individual determinations that are inconsistent with each other at several sigmas. Therefore no definitive resolution is yet available for $\Omega$, or M/L on these large scales; we present the currently available range for comparison purposes only.

The $M/L_B$ ratio in Fig. 2 increases with scale up to the largest observed extent of individual galaxies, R $\sim$ 200 kpc. At this radius, the measured $M/L_B$ of spirals is $\sim$ 100h. Isolated ellipticals have been measured so far up to $\sim 100h^{-1}$ kpc, where $M/L_B \sim 200h$. If we assume that elliptical halos have comparable scale to spirals ($\sim$ 200 kpc) then their $M/L_B$ is $\sim 400h$. This may be supported by the few small x-ray groups of galaxies that include only ellipticals and extend to $\sim 200h^{-1}$ kpc, such as HCG62, the x-ray group with a radius of R $\simeq$ 200 kpc; these groups are consistent with the $M/L_B$ line of ellipticals to R $\sim 200h^{-1}$ kpc (Fig. 2). Beyond this scale, $M/L_B$ appears to "flatten" rather than increase with scale;

the observed $M/L_B$ of groups and clusters typically range between $M/L_B \sim 100$ to $400h$, as expected for a mix of spirals and ellipticals.

## 3. Discussion

We suggest that the total mass of groups, clusters, and large superclusters can be accounted for by their member galaxies plus the hot intracluster gas (observed to account for $\sim$ 5-10% $h^{-1.5}$ of the total mass; Jones & Forman 1992; Bohringer 1994 and references therein). The extended dark halos of galaxies may be stripped-off in dense cluster environments but still remain in the clusters. Since groups, clusters, and superclusters are all composed of a mix of spiral and elliptical galaxies, we expect that their $M/L_B$ (for a system size $R \gtrsim 0.2h^{-1}$ Mpc) will typically range from $\sim$ 100h (if most members are spirals) to $\sim$ 400 h (if most are ellipticals). Under this scenario, the high $M/L_B$ of rich clusters is mainly caused by the high fraction of ellipticals in the clusters [which contain up to $\sim$ 80% ellipticals and SOs (Oemler 1974; Postman & Geller 1984)]. For small groups with $R < 0.2h^{-1}$ Mpc we find intermediate $M/L_B$ ratios as expected for a mix of spirals and ellipticals at that scale (Fig. 2), further supporting the above picture.

On larger scales, the supercluster data is less certain but also suggests that the $M/L_B$ ratio does not continue to rise significantly to large scales; rather, a constant asymptotic value that is consistent with a mixture of spirals and ellipticals may be suggested. The bulk-flow and anisotropy $\beta$ determinations on very large scales are too uncertain at the present time to help constrain the extension to these scales.

Figure 3 presents the ratio of the total mass of the system (galaxies, groups, clusters, superclusters) to the sum of the masses of all the galaxies in the system plus the mass of the hot intracluster gas. The galaxy masses are calculated using $M/L_B \sim 100h$ for spirals and $\sim 400h$ for ellipticals and the known or approximate mean spiral fraction in the system (Oemler 1974; Postman & Geller 1984); the intracluster gas mass is either taken directly from the data or approximated by the mean value ($\sim$ 10% of the total mass of groups, clusters, and superclusters).

Figure 3 shows that all systems with scales $R \gtrsim 0.2$ Mpc have a mass ratio $M/(\Sigma M_{gal} + M_{gas}) \sim 1$, implying that their mass is mainly made up of the total mass of their member galaxies (including their halo mass) plus the intracluster gas. No additional dark matter is needed to account for the mass of groups, clusters, and possibly superclusters if the spiral and elliptical halos behave as suggested by Figs. 1 and 2.

The mass density of the universe implied by this picture can be estimated from the asymptotic value of $M/L_B \sim 300$ h [or more appropriately, $M/L_B \sim 200$h for a typical 70%: 30% mix of Sp: E/S0 in the universe (Postman & Geller 1984) + $\sim 10$% gas]. Combined with the observed mean luminosity density of the universe (Efstathiou *et al.* 1988) of $\sim 2$ x $10^8$ h $L_\odot$/Mpc$^3$ in these units, this yields a universal mass density of $\Omega \sim 0.15 - 0.2$ (where $M/L_B \sim 1350$h is needed for $\Omega = 1$). The total $\Omega$ may increase to $\sim 0.2 - 0.3$ if a small mass bias factor of $b \sim 1.5$ exists on large scales, as suggested by some cosmological simulations (Cen & Ostriker 1992); such a bias is not excluded by current observations (Fig. 2). If, however, $\Omega = 1$ is established on other grounds, then the dark matter must be considerably smoother (i.e., biased) than the distribution of galaxies, clusters, and possibly superclusters, and should exhibit an increase in $M/L_B(R)$ on large scales

## 4. Conclusions

The main conclusions are summarized below:

1. We suggest that most of the dark matter resides in large galaxy halos with $R \sim 200$ kpc. We show that the mass of groups, clusters, and possibly superclusters of galaxies may be accounted for by the mass of their member galaxies, including their large halos (which may be stripped-off in clusters but remain in the system), and the mass of the observed intracluster gas. No additional dark matter is needed to account for the mass of these large systems. We show that the mass-to-light ratio of galaxy systems increases with radius up to $R \sim 0.2 h^{-1}$ Mpc, but appears to remain approximately constant for scales $\gtrsim 0.2 h^{-1}$ Mpc, reflecting the proper mix of spiral and elliptical galaxies (see below).

2. We suggest that the relatively large observed $M/L_B$ ratio of clusters of galaxies ($\sim 300$h)

results mainly from a high $M/L_B$ ratio for elliptical/SO galaxies. We show that ellipticals may have an $M/L_B$ ratio that is approximately *three to four times* larger than typical spirals at the same radius [$(M/L_B)_e \sim 400h$ and $(M/L_B)_s \sim 100h$ within $R \lesssim 200h^{-1}$ kpc]. Since clusters are dominated by elliptical and SO galaxies, a high $M/L_B$ ratio results.

3. The observed asymptotic $M/L_B$ ratio of $\sim 200 - 300h$ on large scales, combined with the observed luminosity density in the universe, suggests a low mass-density universe: $\Omega \simeq 0.15 - 0.2$, with most of the dark matter residing in large galaxy halos. Including a possible mass bias factor of $b \sim 1.5$ on large scales, the density may increase to $\Omega \sim 0.2 - 0.3$, still consistent with the present data. A low density universe is consistent with other observations including the high baryon fraction in clusters (White *et al.* 1993), various large-scale structure observations (Bahcall & Cen 1992), and, indirectly, the large Hubble constant recently reported from Cepheid observations in Virgo (Freedman *et al.* 1994).

We thank J.J. Goodman, M. Gramann, J.P. Ostriker, B. Paczynski, D.N. Spergel, and S. White for helpful discussions. This work was supported by NSF grant AST93-15368 and NASA graduate-student training grant NGT-51295.

**Figures**

FIG. 1  The mass-to-light ratio of spiral and elliptical galaxies as a function of scale. The large boxes indicate the typical ($\sim 1\sigma$) range of $M/L_B$ for bright ellipticals and spirals at their luminous (Holmberg) radii.  ($L_B$ refers to *total* corrected blue luminosity; see text.)  The best-fit $M/L_B \propto R$ lines are shown.

FIG. 2  A composite mass-to-light ratio of different systems - galaxies, groups, clusters, and superclusters - as a function of scale.  The best-fit $M/L_B \propto R$ lines for spirals and ellipticals (from Fig. 1) are shown.  We present median values at different scales for the large samples of galaxies, groups and clusters, as well as specific values for some individual galaxies, x-ray groups, and superclusters.  Typical $1\sigma$ uncertainties and $1\sigma$ scatter around median values are shown.  Also presented, for comparison, are the $M/L_B$ (or equivalently $\Omega$) determinations from the Cosmic Virial Theorem, the Least Action method, and the *range* of various reported results from the Virgocentric infall and large-scale bulk flows (assuming mass traces light).  The $M/L_B$ expected for $\Omega=1$ and $\Omega = 0.3$ are indicated..

FIG. 3a  The mass ratio of galaxies, groups, clusters and superclusters as a function of scale.  The mass ratio represents the ratio of total mass of the system to the sum of the masses of the member galaxies (including the large halo mass) plus the intracluster gas mass (see text).  The boxes indicate the typical ($\sim 1\sigma$) range for the median and individual data values of Fig. 2 for $R \gtrsim 0.1h^{-1}$ Mpc.

FIG. 3b  Same as Fig. 3a but as a function of total luminosity.

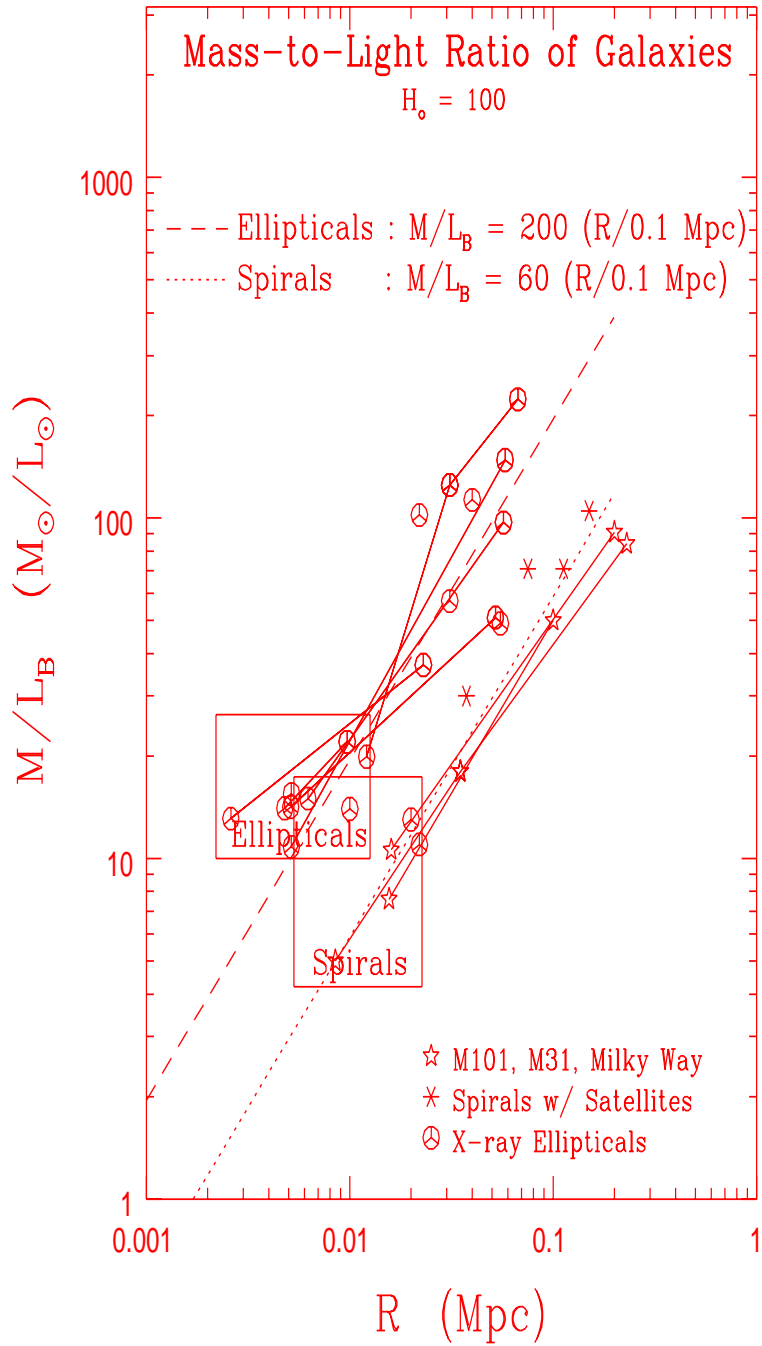

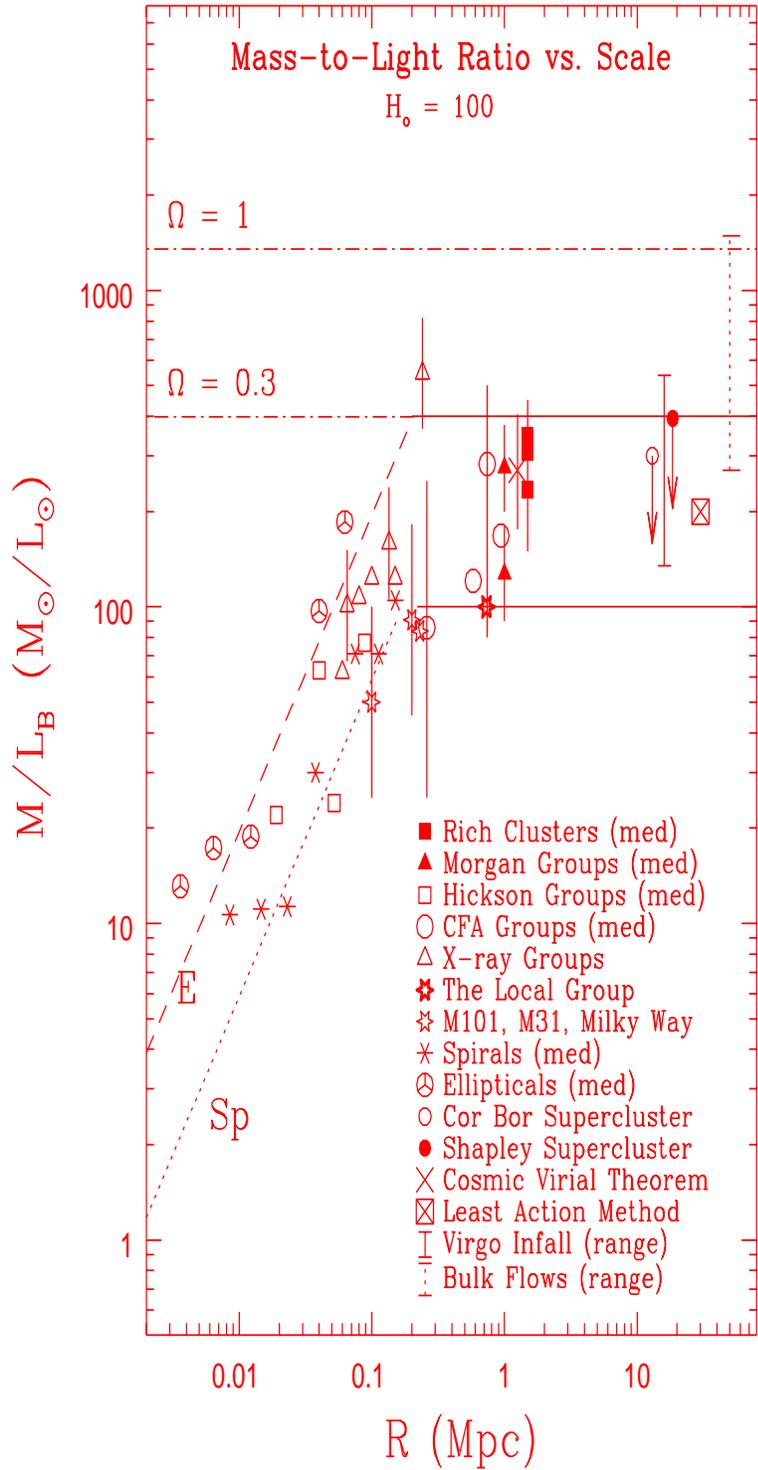

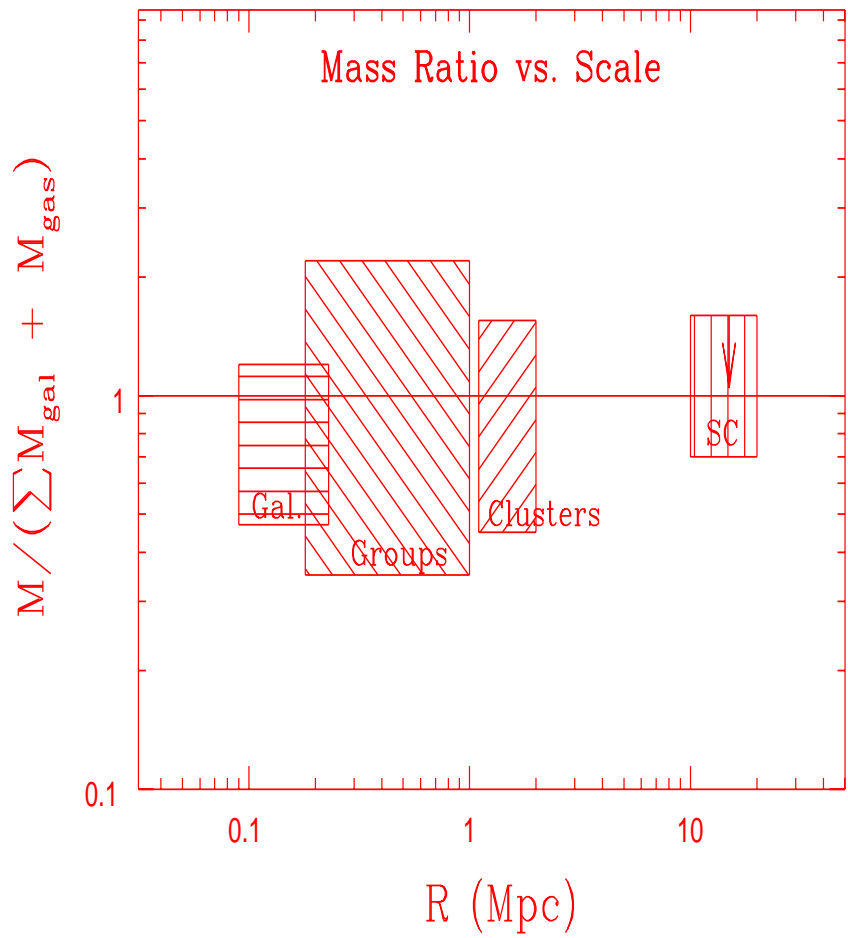

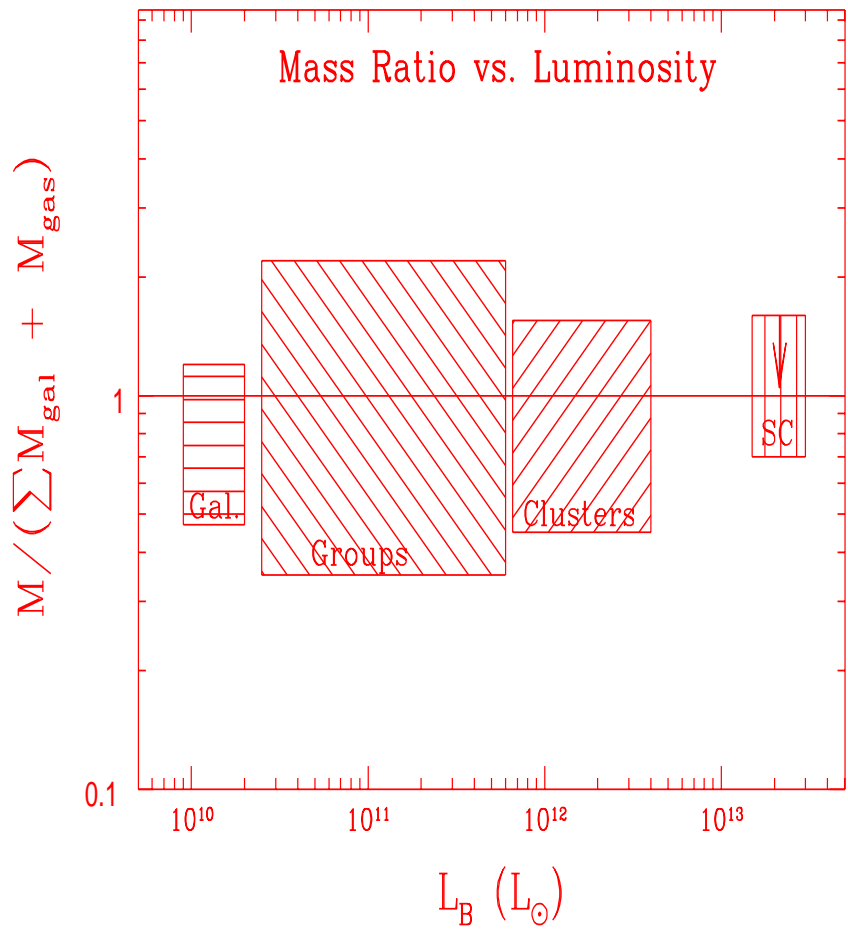